\documentclass[prb,floatfix,nofootinbib]{revtex4}

\usepackage{amssymb}
\usepackage{amsmath}
\usepackage{amsfonts}
\usepackage{graphicx}

\begin{document}

\title{Quantum Malware}
\author{Lian-Ao Wu$^{(1)}$ and Daniel A. Lidar$^{(1,2)}$}
\affiliation{$^{(1)}$Chemical Physics Theory Group, Department of Chemistry, and Center
for Quantum Information and Quantum Control, University of Toronto, 80 St.
George St., Toronto, Ontario, M5S 3H6, Canada\\
$^{(2)}$Departments of Chemistry, Electrical
Engineering-Systems, and Physics, University of Southern California, Los Angeles, CA 90089}

\begin{abstract}
When quantum
communication networks 
proliferate they will likely be subject to a new type of attack: by
hackers, virus makers, and other malicious intruders. Here we introduce the
concept of \textquotedblleft quantum malware\textquotedblright\ to describe
such human-made intrusions. We offer a simple solution for storage of
quantum information in a manner which protects quantum networks from quantum
malware. This solution involves swapping the quantum information at random
times between the network and isolated, distributed ancillas. It applies to
arbitrary attack types, provided the protective operations are themselves
not compromised.
\end{abstract}

\maketitle

\section{Introduction}
  Quantum information processing (QIP) offers unprecedented advantages
compared to its classical counterpart \cite{Nielsen:book}. Quantum
communication is moving from laboratory prototypes into real-life
applications. For example, quantum communication networks (\textquotedblleft
quantum internet\textquotedblright\ \cite{Dowling:03}) have already been
completed, and even commercialized \cite{Elliott:04}. Efforts to protect
quantum information flowing through such networks have so far focused on
environmental (decoherence) and cryptographic (eavesdropping)
\textquotedblleft attacks\textquotedblright . Quantum error correction has
been developed to overcome these disturbances
\cite{Steane:99a,Cleve:99a,Shor:00}. 

Malware (a portmanteau of \textquotedblleft malicious
software\textquotedblright ), familiar from classical information networks,
is any software developed for the purpose of doing harm to a computer system 
\cite{wiki}. This includes self-replicating software such as viruses, worms,
and wabbits; software that collects and sends information, such as Trojan
horses and spyware; software that allows access to the computer system
bypassing the normal authentication procedures, such as backdoors, and more.
In view of their strategic importance, when quantum information networks
become widespread, it is likely that deliberately designed malware will
appear and attempt to disrupt the operation of these networks or their
nodes. We call the quantum version of these types of attacks \emph{quantum
malware}. 

Quantum malware is a new category of attacks on quantum information
processors. While it shares the \textquotedblleft intelligent
design\textquotedblright\ aspect of eavesdropping in quantum cryptography,
one cannot assume that its perpetrators will attempt to minimally disturb a
QIP task. Instead, while quantum malware will try to remain hidden until its
scheduled launch, its attack can be strong and deliberately destructive.
Moreover, generally it should be assumed that malware is able to attack at
any point in time and target any component and part of the quantum devices
in a quantum network. Quantum malware may appear in the form of a quantum
logic gate, or even as a whole quantum algorithm designed and controlled by
the attackers. In comparison with classical information processing, there
are more ways to attack in QIP, because quantum states contain more degrees
of freedom than their classical counterparts. 

Here we
propose a simple scheme to protect quantum memory in quantum information
processors against a wide class of such malware. This scheme, while not
foolproof, dramatically reduces the probability of success of an attack,
under reasonable assumptions, which involve strengthening the defenders
relative to the attackers. We note that if attacker and defender have
exactly the same capabilities (including knowledge, e.g., of secret keys), a
defense is likely to be impossible. Therefore, the question becomes, how
much must one add to the defenders' capabilities, or subtract from the
attackers', in order to have a secure network? The protocol we propose here
to defend against quantum malware provides a possible answer to this
question.

\section{Can quantum malware exist?}
An early no-go theorem showed that it is not possible to build a fixed,
general purpose quantum computer which can be programmed to perform an
arbitrary quantum computation \cite{Nielsen:97a} . However, it is possible
to encode quantum dynamics in the state of a quantum system, in such a way
that the system can be used to stochastically perform, at a later time, the
stored transformation on some other quantum system. Moreover, this can be
done in a manner such that the probability of failure decreases
exponentially with the number of qubits that store the transformation \cite{Vidal:02}. Such stochastic quantum programs can further be used to perform
quantum measurements \cite{rosko:062302,fiurasek:032302,d'ariano:090401}. Thus
it is entirely conceivable that quantum malware can be sent across a quantum
information network, stored in the state of one or more of the network
nodes, and then (stochastically) execute a quantum program or measurement.
Either one of these eventualities can be catastrophic for the network or its
nodes. In the case of a maliciously executed measurement the outcome can be
an erasure of all data. In the case of a quantum program one can imagine any
number of undesirable outcomes, ranging from a hijacking of the network, to
a quantum virus or worm, which replicates itself (probabilistically, due to
the no-cloning theorem \cite{Wootters:82,Dieks:82}) over the network.

\section{Quantum Malware Model and Assumptions}

While there is no limit to the number and character of possible malware
attacks, they must all share the same fundamental characteristic:\ they
comprise a set of elementary operations, \textquotedblleft quantum
machine-language\textquotedblright , such as quantum logic gates and
measurements. It is this simple observation, which also guided the early
concept of the circuit model of quantum computing \cite{Deutsch:89}, that
allows us to consider a general model of quantum malware, without resorting
to specific modes of attack. We thus model quantum malware at this machine-language level.
Clearly, this captures all \textquotedblleft high-level\textquotedblright\
types of attack, since these must, by necessity, comprise such elementary
operations. The operations can be unitary gates $U(t)$ driven by a
time-dependent Hamiltonian $H(t)$, and/or measurements, taking place while a
QIP task is in progress. We denote the series of malicious operations by the
superoperator $\tilde{M}(\{|i\rangle \langle j|\}^{\otimes K})$, where $%
|i\rangle $ is an arbitrary basis state in the Hilbert space in which a
qubit (one of $K$) is embedded. This notation includes measurements, as well
as \textquotedblleft leakage\textquotedblright\ operations that couple the
two states of any qubit with the rest of its Hilbert space. For example, $%
\tilde{M}(\{|i\rangle \langle j|\})$ may have the structure of a quantum
completely positive map \cite{Kraus:83}. This captures the most general type
of quantum malware possible. The details of such quantum malware operations,
i.e., the structure of $\tilde{M}(\{|i\rangle \langle j|\})$, are in general
known only to the attackers, and we will not presume or need any such
knowledge.

In order to protect against malware, in classical information processing one
must assume that there is a means by which to determine, or at least
estimate, a time interval $\delta $ within which the malware is off, so that
malware-free data can be copied (backed up). For example, when one installs
a firewall, or when one applies an anti-virus program, one must assume that
these tasks themselves are malware-free. Similarly, we will assume that the
quantum malware attack occurs in relatively short bursts, and that there are
periods during which there is no attack. We note that it is in the interest
of the attackers to remain hidden, or at least not to launch a continuous
attack. For, otherwise, the defenders may simply decide that it is too
risky to engage in any kind of activity, thus defeating the purpose of the
attackers.

\section{Network operations protocol}

Although the classical backup method is not directly applicable in the case
of quantum information -- because of the no-cloning theorem
\cite{Wootters:82,Dieks:82} -- the basic idea of assuming malware-off
periods 
while copying suggests an analogous mechanism for protecting quantum
information against quantum malware. We note that the assumption that the
attack is switched off every once in a while is not only reasonable for the
sake of the adversary's purpose of maintaining an element of surprise, but
is common also to quantum cryptography. For example, a probabilistic
protocol for quantum message authentication (essentially a \textquotedblleft
secure quantum virtual private network\textquotedblright ) assumes that the
sender and receiver are not subject to attacks by a third party at least
while sending and measuring quantum states \cite{Barnum:02}.

The protocol we describe below is deterministic and is designed to protect
quantum information over time. The networks we consider comprise $K$ nodes,
which can either be the whole network or a part thereof. Each node contains
a quantum computer. The network is used for the transmission of quantum
information. Hence the nodes are connected via quantum and classical
channels. The quantum channels are used for tasks requiring the transmission
of quantum states, such as quantum cryptography \cite{Gottesman:00}. The
classical channels are useful, among other things for teleportation \cite%
{Bennett:93}. Henceforth, the terms \textquotedblleft
online\textquotedblright /\textquotedblleft offline\textquotedblright\
applied to a network node mean that this node is connected/unconnected to
the network. The defenders have access to three types of qubits, or quantum
computers:\ (i) \emph{Data qubits}, which can be either online or offline;
(ii) \emph{Decoy qubits}, which are online when the data qubits are offline,
and vice versa; (iii) \emph{Ancilla qubits}, which are always offline. In
Table~1 we compare the assumptions we make about the respective
capabilities of the defenders and attackers of the network. With the
exception of the limitations listed in Table~1, the attackers are
bound only by the laws of physics. Both defenders and attackers have access
to clock synchronization,\cite{Giovannetti:01}, which enables the defenders
to make use of their set of secret network on-times.

One can envision any number of different methods by means of which the task
of secure distribution of the network on-times to the defenders can be
accomplished, including classical \cite{Blakely:79,Shamir:79} and quantum secret
sharing protocols \cite{Hillery:99,Karlsson:99,Gottesman:00a}, which are
procedures for splitting a message into several parts so that no subset of
parts is sufficient to read the message, but the entire set is. It is
essential to the success of the protocol that only trusted parties are
recipients. The secret set $\{T_{i}\}$ is stored off-line by the defenders,
and is never copied onto a computer that is accessed by the network. This
provision is meant to preclude the attackers from ever gaining access to the
times $\{T_{i}\}$, even if at some point they successfully (remotely) hijack
a network node.

\begin{figure}[th]
  \centering {\includegraphics[angle=270,scale=0.8]{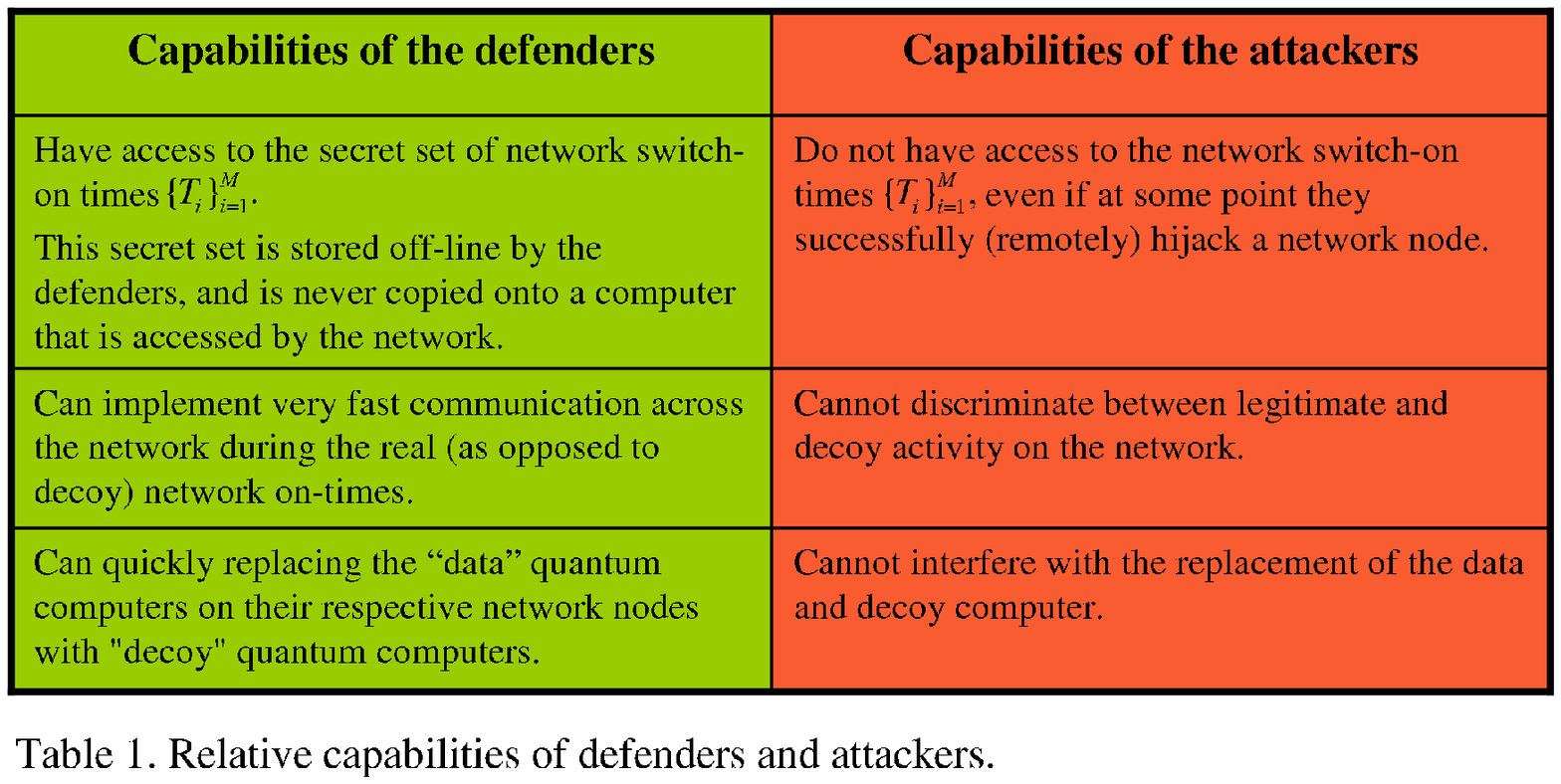}}
  \vspace{-6cm}
\end{figure}

\begin{figure}[th]
\centering {\includegraphics[angle=0,scale=0.9]{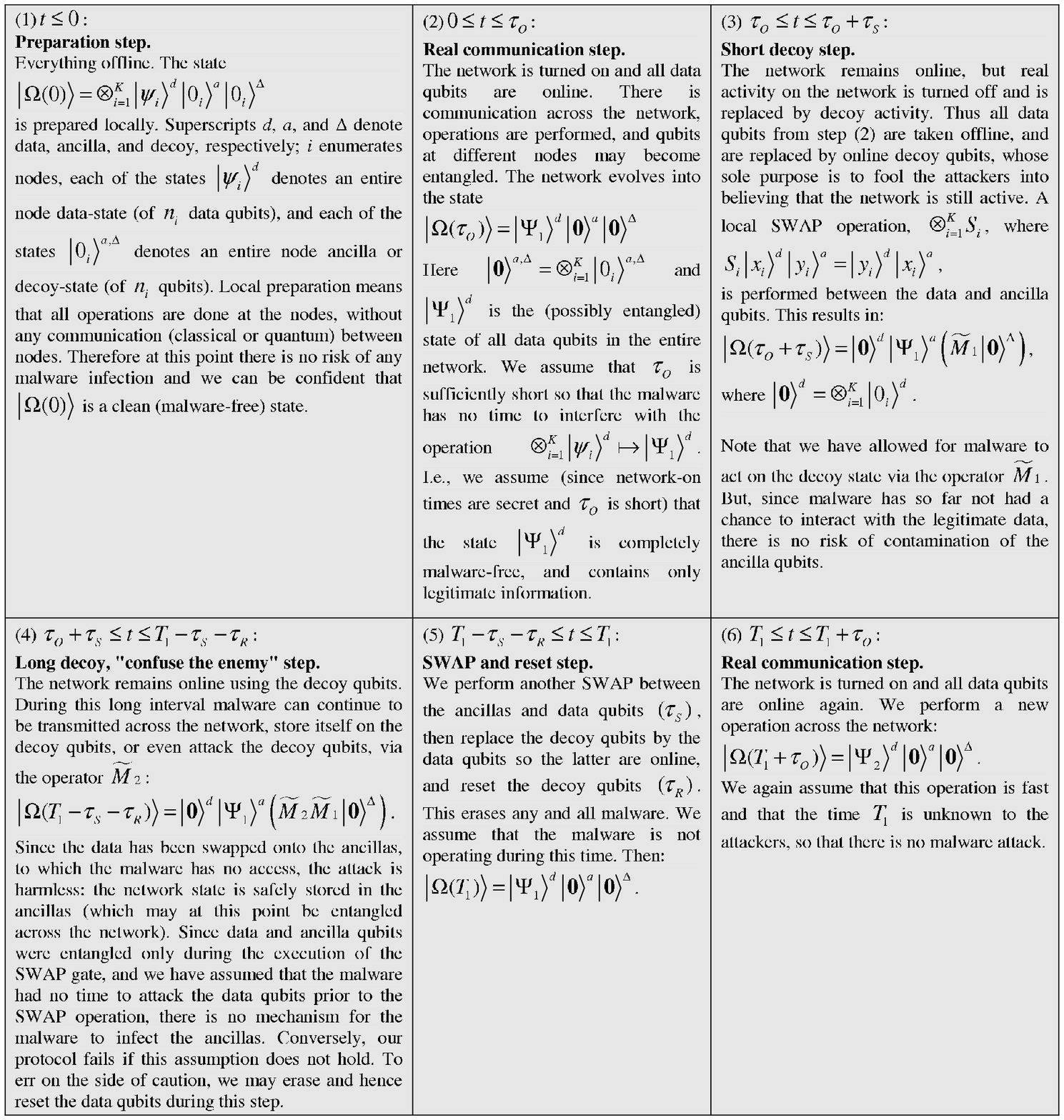}}
\caption{Network operations protocol.}
\label{fig-protocol}
\end{figure}

\begin{figure}[th]
\centering {\includegraphics[angle=270,scale=0.7]{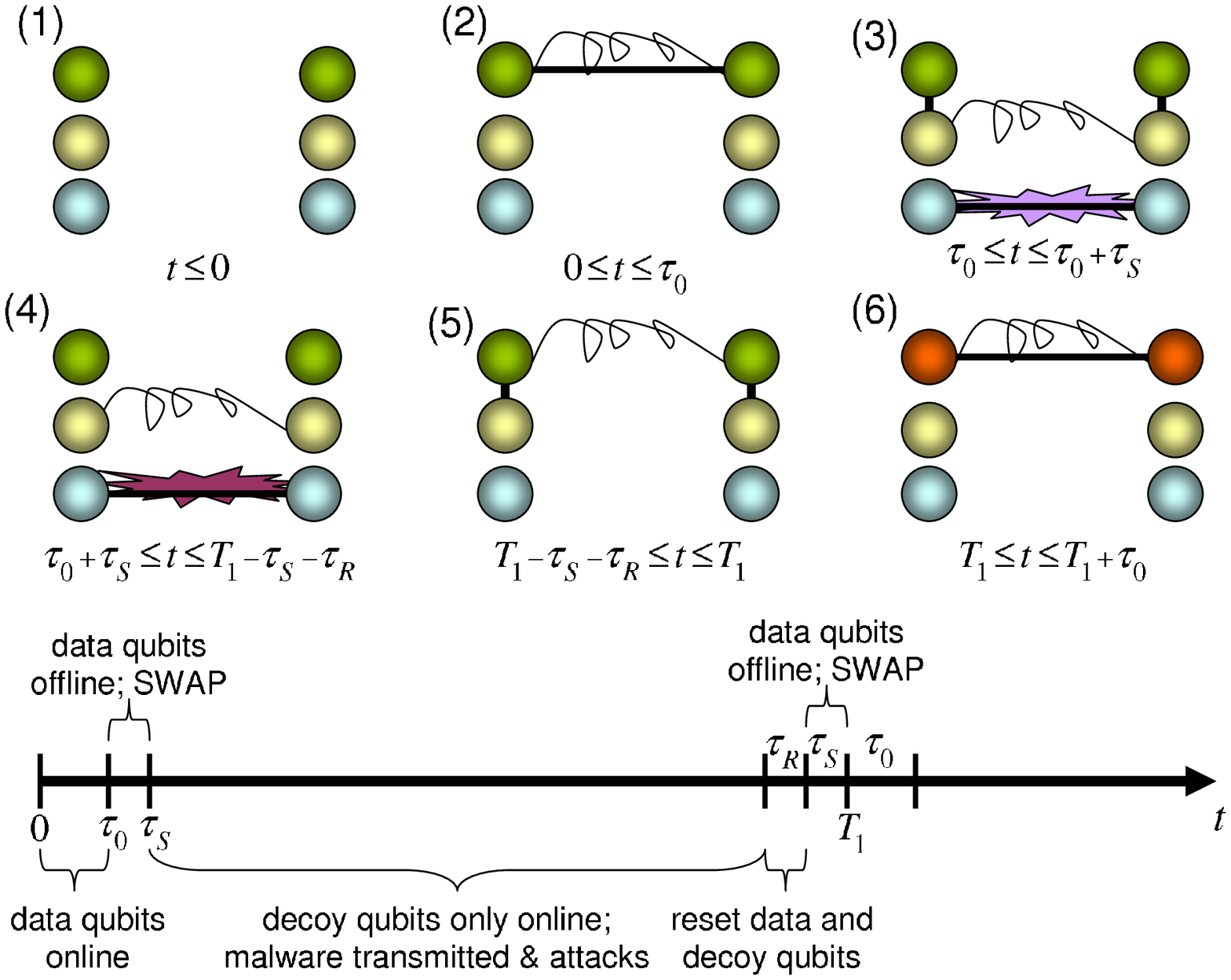}}
\caption{Schematic of network operations protocol. Depicted in the top six parts is a
simple network with $K=2$ nodes and one data qubit in each node. For
simplicity we do not depict other parts of quantum computers where the
malware would reside. Parts (1)-(6) denote the first six steps in the
protocol, starting from the first network cycle. The green dots (top) are
data qubits, the yellow dots (middle) are ancillas, and the blue dots
(bottom) are decoy qubits. Initially (1), the system is offline. When data
qubits are connected by straight lines (2), the system is online. The curly
lines (2) represent entangled qubits. The time at which the network is
turned on is random and unknown to the malware makers, and the duration too
short for them to interfere. In the ultrashort step (3) the network is off
and the state of data and ancilla qubits is swapped, as represented by the
vertical straight lines. The decoy qubits may be under attack. (4) Decoy
qubits are subject to a malware attack. Whatever the attack, in (5) the data
and decoy qubits are reset and the data qubits swapped with the ancilla
qubits. Red data qubits (6) indicate the end of a network cycle, and the
start of a new cycle. Bottom part: Timeline of the protocol.}
\label{f1}
\end{figure}

Our network protection protocol is given in Figure~\ref{fig-protocol}. After
the preparatory steps (1) and (2), the protocol cycles through steps
(3)-(6), with the next network on-times $\{T_{i}\}$ chosen from the
previously distributed secret set. The protocol is further illustrated in
Figure~\ref{f1}.

\section{Conditions for success of the defense protocol}

We note that if a malware attack ever takes place during the
network-on times, replacement of data qubits by decoy qubits, decoy-reset,
or SWAP operations, the protocol fails and the network must be completely
reset. We can estimate the probability, $p$, of this catastrophic occurrence
as follows. A reasonable strategy is to pick the times $\{T_{i}\}$ from a
uniformly random distribution. The malware designers, on the other hand, may
choose their attack interval times $\{\theta _{j}\}$ from some other
distribution, not known to us. Let us characterize this latter distribution
by a mean attack interval $\theta $ and mean attack length $\delta $. Let
the total time over which the protocol above is implemented be $T$. Let us
also designate the operating times within a single cycle of our protocol by $%
\tau $ (i.e., $\tau =\tau _{O}+2\tau _{S}+\tau _{R}$). Consider a particular
attack window of length $\delta $ at some random time. The probability $%
q_{1} $ that the network is \emph{off} during this window is $%
q_{1}=[T-(\delta +\tau )]/T$, since there are two network-on intervals, one
before and one after the attack window, and each must be a distance $\tau /2$
away from this window. In other words, the excluded interval is $\delta
+2(\tau /2)$. Now, since the network-on times are randomly distributed, the
probability that this same attack window does not overlap any network-on
interval, after $M$ such intervals, is $q_{M}\approx q_{1}^{M}$ (this is an
approximation since one should actually exclude overlapping intervals \cite%
{Hamburger:random-model}, but if the intervals are sufficiently sparse such
overlaps can be neglected). The probability of at least one (catastrophic)
overlap of this attack window with a network-on interval is $p=1-q_{M}$.
Letting $M=cT$, where $0<c<1$ is a constant, we have $p\overset{T\rightarrow
\infty }{\longrightarrow }1-\exp [-c(\delta +\tau )]$, so that as long as $c$
and $\tau $ (under our control), and $\delta $ (under the attackers'
control) are sufficiently small, we have $p\approx c(\delta +\tau )\gtrsim 0$%
. Another way of analyzing the optimal strategy is to note that there are,
on average, a total of $A=T/(\theta +\delta )$ attack intervals, so that the
expected number of catastrophic overlaps is $Ap=[T/(\theta +\delta
)]\{1-[1-(\delta +\tau )/T]^{M}\}$, and this number must be $\ll 1$ for our
protocol to succeed. Given an estimate of the attackers' parameters, $\theta 
$ and $\delta $, and given that the state of technology will impose a
minimum $\tau $, we can use this result to optimize $T$ and $M$. A simple
estimate can be derived in the physically plausible limit $\delta ,\tau \ll
T $, where we can linearize the above expression and obtain the condition 
\begin{equation}
M\ll (\delta +\theta )/(\delta +\tau ).
\end{equation}%
If we further assume $\tau <\delta \ll \theta $, we find the intuitively
simple result that {\em the number of network-on times cannot exceed the ratio of
the mean attack interval to the mean attack length}.

We note that one might suspect that our protocol is in fact more
vulnerable than suggested by the arguments above, given that an adversary
might hijack quantum repeaters installed between network nodes and tweak the
data (this scenario assumes quantum optical communication). However, we
point out that there exists an alternative scheme to the use of quantum
repeaters:\ in order to overcome photon decoherence and loss one may use a
spatial analog of the quantum Zeno effect and \textquotedblleft bang-bang
decoupling\textquotedblright , which involves only linear optical elements
installed at regular intervals along an optical fiber \cite{WuLidar:04}.
Such a system cannot be hijacked because of its distributed nature. An
attacker could at most remove, or tamper with, some of the linear optical
elements, thus degrading the performance of the quantum noise suppression
scheme.

\section{Detection of an attack, and increasing the robustness of the defense protocol}

A considerable improvement in the robustness of the stored
quantum information is possible by replacing the SWAP operation with an
encoding of each data qubit into a quantum error-detecting code \cite{Nielsen:book}. Not only
does this enable the application of quantum fault tolerance methods
\cite{Steane:99a}, {\em it also allows the defenders to check whether the
data has been modified}, via the use of quantum error detection. However, since this
does not allow us to change our assumptions about the relative
weakness/strength of attackers and defenders, we do not here consider this
possibility in detail. 

We further note that it is possible to slightly relax the assumption that
the malware makers cannot interfere during the real communication step (2).
Indeed, it is possible to let the malware attack and/or store itself on
another set of qubits connected to the network, as long as these qubits are
not involved in storing the legitimate state being processed across the
network. When executing the short decoy step (3), we must then assume that
this other set of qubits does not interact with the ancillas.

\section{Implementation of SWAP gates}
We now show how the SWAP gates needed in our protocol can be implemented
in a variety of physical systems. Recall that above we distinguished between
malware operating on the qubits' Hilbert space, and malware that includes
operations on a larger Hilbert space (\textquotedblleft
leakage\textquotedblright ). The implementation of $S_{i}$ for malware $%
\tilde{M}(\left\{ \vec{\sigma}_{i_{d}}\right\} )$, where $\vec{\sigma}_{i_{d}}$ are the
Pauli matrices on the data qubits $i_d$, without leakage, is
direct. Assume that the Heisenberg interaction $\vec{\sigma}_{i_{d}}\cdot 
\vec{\sigma}_{i_{a}}$ between the $i$th data qubit and its ancilla is
experimentally controllable, as it is in a variety of solid-state quantum
computing proposals such as quantum dots \cite{Loss:98}. Then the SWAP gate
is $S_{i}=\exp (i\frac{\pi }{2}P_{i_{d}i_{a}})$, where $P_{i_{d}i_{a}}=\frac{%
1}{2}(\vec{\sigma}_{i_{d}}\cdot \vec{\sigma} _{i_{a}}+1)=$ $\sum_{\alpha
,\beta =0}^{1}(\left\vert \alpha \right\rangle _{i_{d}}\left\langle \beta
\right\vert )\otimes (\left\vert \beta \right\rangle _{i_{a}}\left\langle
\alpha \right\vert )$ is an operator exchanging between the $i$th data qubit
and its ancilla, and the gate time $\tau _{\alpha }$ ($\alpha =O,S,R$) is on
the order of a few picoseconds \cite{Burkard:99}. The SWAP gate can be
implemented in a variety of other systems, with other Hamiltonians, in
particular Hamiltonians of lower symmetry \cite{LidarWu:01,Bonesteel:01}.

{If the attackers design malware capable of causing leakage into or
from the larger Hilbert space with dimension $N$, the situation will be
different. Generally, the malware superoperator $\tilde{M}$ is a function of
transition operators of the form $\left\vert \alpha \right\rangle
_{i}\left\langle \beta \right\vert $, where the case of $\alpha ,\beta >1$
represents states other than the two qubit states $|0\rangle $ and $%
|1\rangle $. If both $\alpha $ and $\beta $ are $0$ or $1$, the operation
can be expressed in terms of Pauli matrices, e.g., $\left\vert
0\right\rangle \left\langle 1\right\vert =\sigma ^{x}+i\sigma ^{y}$; if only
one of either $\alpha $ or $\beta $ is $0$ or $1$, the operation represents
leakage to or from the qubit subspace. Let us define a generalized
data-ancilla exchange operator, $P_{i_{d}i_{a}}=$ $\sum_{\alpha ,\beta
=0}^{N-1}(\left\vert \alpha \right\rangle _{i_{d}}\left\langle \beta
\right\vert )\otimes (\left\vert \beta \right\rangle _{i_{a}}\left\langle
\alpha \right\vert )$. Then $P_{i_{d}i_{a}}|\alpha \rangle _{i_{d}}|\beta
\rangle _{i_{a}}=|\beta \rangle _{i_{d}}|\alpha \rangle _{i_{a}}$ and $%
P_{i_{d}i_{a}}^{2}=I$, the identity operator. Therefore the generalized SWAP
operator is }
{
\begin{equation}
S_{i}=\exp (i\frac{\pi }{2}P_{_{i_{d}i_{a}}})=iP_{_{i_{d}i_{a}}},
\label{eq2}
\end{equation}
and it follows directly that $S^{\dagger }\tilde{M}(\left\{ i_{d}\right\}
)S=\tilde{M}(\left\{ i_{a}\right\} )$, where $S=\prod_{i}S_{i}^{\dagger }$%
. The exchange operator $P_{i_{d}i_{a}}$ can be implemented as a
controllable two-body Hamiltonian in multi-level systems. }

{For a fermionic system such as excitonic qubits in quantum dots \cite%
{Biolatti:00,Chen:01} or electrons on the surface of liquid helium \cite%
{Platzman:99}, a qubit is defined as $\left\vert 0\right\rangle
=f_{0}^{\dagger }\left\vert \mathrm{vac}\right\rangle $, $\left\vert
1\right\rangle =f_{1}^{\dagger }\left\vert \mathrm{vac}\right\rangle $,
where $f_{0}^{\dagger },f_{1}^{\dagger }$ are fermionic creation operators
and $\left\vert \mathrm{vac}\right\rangle $ is the effective vacuum state
(e.g., the Fermi level). The most general attack uses an operator (a
Hamiltonian or measurement) that can be expressed in terms of $%
F_{i_{d}}\equiv (f_{0,i_{d}}^{\dagger })^{k}(f_{1,i_{d}}^{\dagger
})^{l}(f_{0,i_{d}})^{m}(f_{1,i_{d}})^{n}$ (where $k,l,m,n$ are integers)
acting on the $i$th data qubit. These operators can be shifted to the
corresponding ancilla, via control of a two-body fermionic Hamiltonian.
Namely, $S_{i}^{\dagger }F_{i_{d}}S_{i}=F_{i_{a}}$, where the SWAP\ operator
for the $i$th fermionic particle reads $S_{i}=S_{i}^{0}S_{i}^{1}$, where 
\begin{equation*}
S_{i}^{q}=\exp [\frac{\pi }{2}(f_{q,i_{d}}^{\dagger
}f_{q,i_{a}}-f_{q,i_{a}}^{\dagger }f_{q,i_{d}})],
\end{equation*}%
with $q=0,1$. This can be proven easily using the identities 
\begin{eqnarray*}
e^{-\phi (f_{d}^{\dagger }f_{a}-f_{a}^{\dagger }f_{d})}f_{d}^{\dagger
}e^{\phi (f_{d}^{\dagger }f_{a}-f_{a}^{\dagger }f_{d})} &=&\cos \phi
f_{d}^{\dagger }+\sin \phi f_{a}^{\dagger }, \\
e^{-\phi (f_{d}^{\dagger }f_{a}-f_{a}^{\dagger }f_{d})}f_{d}e^{\phi
(f_{d}^{\dagger }f_{a}-f_{a}^{\dagger }f_{d})} &=&\cos \phi f_{d}+\sin \phi
f_{a},
\end{eqnarray*}%
which follow from the Baker-Hausdorff formula $e^{-\alpha A}Be^{\alpha
A}=B-\alpha \lbrack A,B]+\frac{\alpha ^{2}}{2!}[A,[A,B]]-...$. The relation $%
S_{i}^{\dagger }F_{i_{d}}S_{i}=F_{i_{a}}$ implies that the action of any
\textquotedblleft fermionic malware\textquotedblright\ is shifted by the
SWAP\ gate from the data to the ancilla particle. The very same construction
works also for bosonic systems, such as the linear-optics quantum computing
proposal \cite{Knill:00}. There a qubit is defined as $\left\vert
0\right\rangle =b_{0}^{\dagger }\left\vert \mathrm{vac}\right\rangle $, $%
\left\vert 1\right\rangle =b_{1}^{\dagger }\left\vert \mathrm{vac}%
\right\rangle $, where $b_{0}^{\dagger },b_{1}^{\dagger }$ are bosonic
creation operators. The relations we have just presented for fermions hold
also for bosons, provided one everywhere substitutes bosonic operators in
place of the fermionic ones.

\section{Conclusions}

What sets quantum malware apart from the environmental and eavesdropping
attacks is that the latter are typically weak (in the sense of coupling to
the quantum information processing (QIP) device), while the former can be
arbitrarily strong, can attack at anytime, and can target any part of a
quantum device. Indeed, a malicious intruder, intent on disrupting
information flow or storage on a quantum network, will resort to whatever
means available. In contrast, the QIP-environment interaction will be a
priori reduced to a minimal level, and an eavesdropper will attempt to go
unnoticed by the communicating parties. For this reason one cannot expect
quantum error correction to be of use against quantum malware, as it is
designed to deal with small errors. The same holds true for quantum
dynamical decoupling \cite{Viola:04} or other types of Zeno-effect like
interventions \cite{Facchi:05}. Decoherence-free subspaces and subsystems 
\cite{LidarWhaley:03}, on the other hand, do not assume small coupling, but
do assume a symmetric interaction, which is unlikely to be a good assumption
in the case of quantum malware. We further note that of all possible types of quantum
malware, as far as we know only quantum trojan horses have been considered
previously, in the quantum cryptography literature. In particular, in the
context of the security proof of quantum key distribution, it was shown that
teleportation can be used to reduce a quantum trojan horse attack to a
classical one \cite{Lo:99}. Finally, we note that the attacks we are concerned with are
on the quantum data, not the quantum computer software; the latter is
generally itself a list of classical instructions, and can be cloned.

Experience with classical information processing leaves no doubt that the
arrival of quantum malware -- malware designed to disrupt or destroy the
operation of quantum communication networks and their nodes (quantum
computers) -- is a matter of time. When this happens, overcoming the problem
of quantum malware may become as important as that of overcoming
environment-induced decoherence errors. In this work we have raised this
specter, and have offered a relatively simple solution. Our solution invokes
a network communication protocol, wherein trusted parties operate the
network at pre-specified times, and quickly swap the information out of the
network onto a quantum backup system. Such a protocol slows the network down
by a constant factor, and therefore does not interfere with any quantum
computational speedup that depends on scaling with input size. The success
of our protocol depends strongly on the ability to perform very rapid
swapping between data and ancilla qubits. This suggests the importance of
the design of fast and reliable swapping devices. This
can be done for a variety of physical systems, as shown above. As long as the swapping can be done sufficiently fast, and as long
as there exists a mechanism for secure distribution of the network on-times
only among trusted parties, we have shown that the quantum network will be
unharmed by a very general model of quantum malware. On the other hand, if
these assumptions are not satisfied and an attack is successful, one must
unfortunately reset the network, pending the development of a
\textquotedblleft quantum anti-virus program\textquotedblright\ that would
clean infected data. The latter is a very interesting open research problem.
Our protocol is similar to \textquotedblleft paranoid\textquotedblright\
classical protocols employed in military systems that are under attack,
which are shut down a great deal of the time, and then are suddenly opened
up in order to perform a useful task. However, there is a distinct quantum
aspect to our protocol, which is that it preserves entanglement across the
network. In this sense our protocol, while being a conceptually simple
generalization of established classical methods, offers a genuine step
forward towards quantum network security against quantum malware.

\begin{acknowledgments}
Financial support from the DARPA-QuIST
program (managed by AFOSR under agreement No. F49620-01-1-0468) and the
Sloan Foundation (to D.A.L.) is gratefully acknowledged. We thank Dr. Y. Xu
(Microsoft Research Asia, Beijing) and Prof. Hoi-Kwong Lo (University of
Toronto) for helpful discussions.
\end{acknowledgments}


\end{document}